# Preprint: Bringing immersive enjoyment to hyperbaric oxygen chamber users using virtual reality glasses

Zhihan Lv
FIVAN, Valencia, Spain
lvzhihan@gmail.com

## ABSTRACT
This is the preprint version of our paper on REHAB2015. This paper proposed a novel immersive entertainment system for the users of hyperbaric oxygen therapy chamber. The system is a hybrid of hardware and software, the scheme is described in this paper. The hardware is combined by a HMD (i.e. virtual reality glasses shell), a smartphone and a waterproof bag. The software is able to transfer the stereoscopic images of the 3D game to the screen of the smartphone synchronously. The comparison and selection of the hardware are discussed according to the practical running scene of the clinical hyperbaric oxygen treatment. Finally, a preliminary guideline for designing this kind of system is raised accordingly.

## Categories and Subject Descriptors
H.4 [**Information Systems Applications**]: Miscellaneous; D.2.8 [**Software Engineering**]: Metrics—*complexity measures, performance measures*

## General Terms
Theory

## 1. INTRODUCTION
The efficacy of hyperbaric oxygen therapy has been recorded in a lot of literature in clinical research community [30] [26] [9] [27]. The limited space of the hyperbaric oxygen chamber has constrained the possibilities of activities. The most common activity that the users usually do in the hyperbaric oxygen chamber is sleeping, although the oxygen generation machine of the chamber is noisy which couldn't be ignored at all. Playing with smartphone or reading book also happens sometime, since they only need activities of upper arms. It's however still not comfortable to retain the postures of the upper arms and neck to play or read for long time. The physical discomforts may impair the psychological enjoyment.

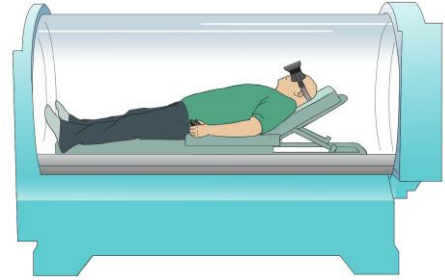

**Figure 1: The system running scenarios.**

Overall, the spatial limitation constrains the physical activities inside the hyperbaric oxygen chamber. Our research plans to improve the user experience of hyperbaric oxygen therapy by novel virtual reality (VR) based interactive technologies, the imagination of the running scenario is shown as in figure 1.

Virtual reality (VR) environments are increasingly being used by neuroscientists [2] and psychotherapists [28] to simulate natural events and social interactions. VR technology has been proved to be able to stimuli for patients who has difficulty in imagining scenes and/ or are too phobic to experience real situations since long time ago [24]. The 'Immersive' characteristic of VR technology can substantially improve movement training for neurorehabilitation [22] [5] [14] [1]. The report of the early research of utilizing VR to treat claustrophobia patient [3] has inspired our work, since the hyperbaric oxygen chamber is a typical sealed space. The underwater VR game for aquatic rehabilitation brings us some suggestions and tips [13] about the development of VR game on head mounted device (HMD) in unusual air environment (e.g. under water, in hyperbaric oxygen). Nonetheless, VR in clinical rehabilitation is still in infancy and need more exploration [25].

## 2. SYSTEM
In this system, the software could duplicate the stereoscopic images from game window to mobile phone screen by Wifi or USB. So, the 3D games or films could be easily shown on the screen of smartphone without porting the programme to smartphone, as long as there is a PC or laptop nearby as a server.

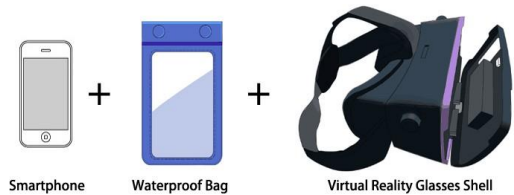

**Figure 2: Waterproof bag is employed to separate the phone to the high-density oxygen.**

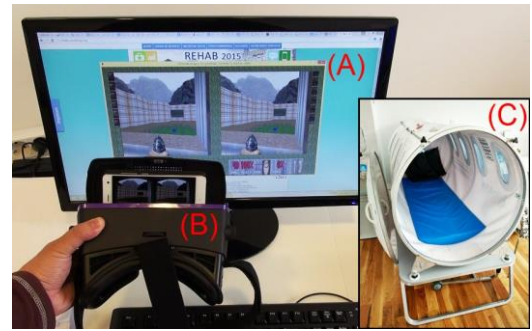

**Figure 3: The graphical content of screen (A) is duplicated to smartphone screen (B). (C) is the hyperbaric oxygen chamber located in our clinic.**

## 2.1 Hardware

The VR hardware has experienced several-generation development and emerged a number of useful immersive environment, which ranges from early surround-screen projection-based VR environment so-called CAVE [7] [6] to latest human-computer-interface (HCI) such as VR glasses. In our case, CAVE is impossible to be taken into the hyperbaric oxygen chamber due to the high cost and the dependence on plane context. Therefore, VR glasses become the best choice for bringing the 3D immersive perception to users lying inside hyperbaric oxygen chamber.

The size of the hyperbaric oxygen chamber is 185CM length, 90CM diameter, in which an adult cannot sit up. The chamber is built by plastic sheeting for insulation and has four small windows on each side, as shown in figure 3 (C). Since the electronic device is not allowed to directly use in the hyperbaric oxygen chamber in order to prevent the explosion, the smartphone has to be sealed into a insulation bag which could separate the phone to the high-density oxygen. In our designed system, we employ a waterproof bag to wrap the smartphone and then put it into the HMD, as shown in figure 2. The smartphone has Quad-Core 2.5GHz CPU, 3GB DDR3 memory, the screen is 1920*1080 resolution, 441PPI, 95% NTSC color gamut, it supports Wifi and bluetooth. The HMD shells we have chosen include four productions as shown in Figure 4.

## 2.2 Software

The smartphone selected in our system runs Android operating system. Other smartphone operating systems are also available in future development.

Two kinds of software technologies can implement the stereoscopic 3D applications on smartphone screen, as described as followed. Virtual Desktop. As shown in Figure 3, the 3D game GZ3DOOM which is the modified version of DOOM is running in stereoscopic 3D mode and the game window is duplicated to the smartphone screen by a screen-synchronization software. The connection between the PC and smartphone is by Wifi. In our system, we employed Trinus VR [34] as the screen-synchronization software, which includes a .exe program running on windows as server side and a .apk application on smartphone as client side. Some other similar software can be accessed from google play too, such as Intugame VR [10]. It worth to mention that, TrinusVR has fake3D function which can generate fake stereoscopic image by automatically duplicating the original image for each eye. Even though the fake3D actually brings flat perception for user, there are also tools that make the conversion from monoscopic of 3D game to stereoscopic side by side (SBS), like Vireio [23] or TriDef 3D [31]. Mobile Phone APP. The 3D game or VR scene application running on smartphone may suffer worse system performance. The available applications resources are not sufficient too since the accumulated 3D games on PC or other game devices in past cannot directly run on smartphone or be simply modified to be suitable for smartphone operating system (i.e. Android). Anyway, the customized mobile APP developed for smartphone has better compatibility with the hardware and probably will be improved by next-generation Cloud operating system such as Windows 10, which is a promising development trend of operating system. The VR glasses SDK (i.e. Cardboard) for latest version of video game engines (i.e. Unity3D) have been provided to adapt existing Unity3D APPs for virtual reality [11], which is a available choice with friendly user interface for non-programmer.

## 2.3 Running Scenarios

The user (e.g. patient, athlete, healthy people) is lying inside the hyperbaric oxygen chamber, wears the HMD and holds the remote controller as shown in figure 1. The user can play the 3D game in stereoscopic 3D mode or 2D game in fake3D mode which can also show the game image on the VR glasses. Moreover, conventional need of reading E-books or accessing Internet are still possible to be achieved on VR glasses by fake3D mode. The input methods include head motion and remote controller. The head motion on VR glasses only supports the rotations around three axis which are controlled by gyroscope sensor of the smartphone. The head rotation actions are synchronous with the rotation of the avatar's view in the VR scene, which brings the immersive perception to the user in real time. Meanwhile, the remote controller is used to input the displacement of the avatar as well as manipulate the menu of the game configuration.

## 3. PRELIMINARY COMPARISON

Figure 4 shows the HMD that we have considered and compared for the clinical need. Where (a)(c)(d) are the VR glasses shell by which users could watch the anaglyph 3D scene generated from smartphone screen. While (b) is so-called 'Lazy Glasses', the lens of which reflects the scene toward the below of user's eyes. As shown in figure 5, the user can lie flat and play the smartphone instead of raising the head or forearms. As we have known, (b) is not a VR technology based device, but it has better suitability. The user could read book or play smartphone game by wearing while lifting the neck or raising arms are needless any more (b). Moreover, as a non-electronic device, (b) is more safe and convenient in hyperbaric oxygen chamber. However, (b) couldn't generate the stereoscopic 3D images for immersive perception which is the essential condition of VR. Anyway, (b) is a convenient choice for users who don't really desire to enjoy 3D immersive experience while still hope to read in the hyperbaric oxygen chamber.

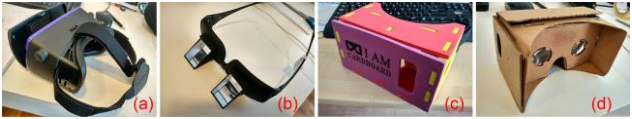

**Figure 4: The HMD that we have considered and compared for the clinical need.**

Comparing (a)(c)(d), the three shells are respectively made by hard plastic, ethylene-vinyl acetate (EVA), and cardboard. The convergence-to-face part of (a) for light blocking is made by soft holster filling of sponge, so it's not oppressive at all. In addition, (a) can modify the pupil distance (PD) and depth of field (DOF), so (a) is suitable for the users with different myopia degree and PD. (c) and (d) have not adequate adaptive functions for users with disparity in physical characteristics, both are however in collapsible structures the portability of which leads to convenience for distribution and transportation. In our practical clinical need, the features of (b) are more suitable for different patients. By the way, the price level of the four devices we have chosen are almost equal, so the cost problems will not affect the motivation of the device choice in future practical application.

The common drawback of the HMDs is the dizziness. Dizziness is mainly caused by the vision delay which leads to vestibular and cochlear imbalance. In our case, lying posture doesn't lead to more dizziness than other postures such as standing and sitting. So we believe the VR glasses will bring immersive enjoyment to hyperbaric oxygen chamber users as long as it can create pleasure to common players.

## 4. DESIGNING GUIDELINE

1. The water proof bag is necessary for isolation from hyperbaric oxygen.
2. It's significant to choose a comfortable HMD shell.
3. The smartphone programming doesn't have to be done as long as a PC or laptop is nearby as a server.
4. An additional remote controller is needed for menu manipulation.
5. Dizziness is not enhanced by hyperbaric oxygen chamber.
6. 'Lazy Glasses' is a non-electrical choice for users who hope to read 2D content.

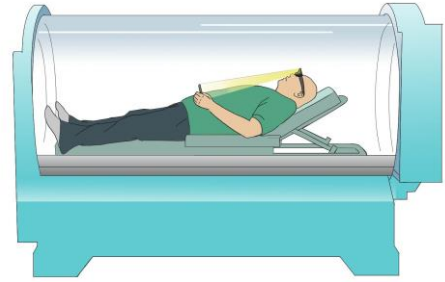

**Figure 5: The user is experiencing 'Lazy Glasses' and playing with mobile phone.**

## 5. CONCLUSION

In this paper, we proposed a hybrid system for improving the user experience in the hyperbaric oxygen chamber. By this system, users can enjoy 3D stereoscopic games wearing virtual reality glasses and immersive perception, as well as read E-books or play the 2D games. Several HMDs have been tested and compared by us. All the technical issues have been solved, the new challenges are the customized 3D games or software used for this case. The efficacy of this system will be evaluated in future research by subjective measurement such as Geneva Emotion Wheel (GEW) [29] as well as the medical scales [3] (e.g. Subjective units of discomfort scale (SUDS), Problem-related impairment questionnaire (PRIQ), Self-efficacy towards the target behaviour measure (SETBM), The attitude towards CTS measure (TAM)). Some novel technology will also be used to improve this research, e.g., Sensors [35], Virtual Rehabilitation [17] [16], HCI [18] [19], Video Game [21] [4], Ubiquitous Computing [20], Control [38], Database [33] [36], Big Data [39] [37], Distributed Computing [32] [15] [12], Optimization Algorithm [8].

### Acknowledgment
The authors would like to thank Pablo Gagliardo and Sonia Blasco for practical clinical suggestions and fruitful discussion about the technical possibility, and thank Dr. Javier Chirivella and Vicente Penades for the help at FIVAN. The authors thank to LanPercept, a Marie Curie Initial Training Network funded through the 7th EU Framework Programme (316748).

## 6. REFERENCES

[1] P. Bajcsy, K. McHenry, H.-J. Na, R. Malik, A. Spencer, S.-K. Lee, R. Kooper, and M. Frogley. Immersive environments for rehabilitation activities. In *Proceedings of the 17th ACM international conference on Multimedia*, pages 829–832. ACM, 2009.

[2] C. J. Bohil, B. Alicea, and F. A. Biocca. Virtual reality in neuroscience research and therapy. *Nature reviews neuroscience*, 12(12):752–762, 2011.

[3] C. Botella, R. Baños, C. Perpina, H. Villa, M. u. Alcaniz, and A. Rey. Virtual reality treatment of claustrophobia: a case report. *Behaviour research and therapy*, 36(2):239–246, 1998.

[4] Z. Chen, W. Huang, and Z. Lv. Towards a face recognition method based on uncorrelated

discriminant sparse preserving projection. *Multimedia Tools and Applications*, pages 1–15, 2015.

[5] L. Connelly, Y. Jia, M. L. Toro, M. E. Stoykov, R. V. Kenyon, and D. G. Kamper. A pneumatic glove and immersive virtual reality environment for hand rehabilitative training after stroke. *Neural Systems and Rehabilitation Engineering, IEEE Transactions on*, 18(5):551–559, 2010.

[6] C. Cruz-Neira, D. J. Sandin, and T. A. DeFanti. Surround-screen projection-based virtual reality: The design and implementation of the cave. In *Proceedings of the 20th Annual Conference on Computer Graphics and Interactive Techniques*, SIGGRAPH '93, pages 135–142, New York, NY, USA, 1993. ACM.

[7] C. Cruz-Neira, D. J. Sandin, T. A. DeFanti, R. V. Kenyon, and J. C. Hart. The cave: Audio visual experience automatic virtual environment. *Commun. ACM*, 35(6):64–72, June 1992.

[8] S. Dang, J. Ju, D. Matthews, X. Feng, and C. Zuo. Efficient solar power heating system based on lenticular condensation. In *Information Science, Electronics and Electrical Engineering (ISEEE), 2014 International Conference on*, volume 2, pages 736–739, April 2014.

[9] J. C. Davis and T. Hunt. Hyperbaric oxygen therapy. *J Intensive Care Med*, 4(2):55–57, 1989.

[10] G. Georgi and N. Krasimir. Intugame vr. http://intugame.com.

[11] Google. Cardboard sdk for unity. https://developers.google.com/cardboard/unity/.

[12] D. Jiang, X. Ying, Y. Han, and Z. Lv. Collaborative multi-hop routing in cognitive wireless networks. *Wireless Personal Communications*, pages 1–23, 2015.

[13] Q. John. Shark punch: A virtual reality game for aquatic rehabilitation. In *Virtual Reality (VR), 2015 iEEE*, pages 265–266. IEEE, 2015.

[14] R. Kizony, N. Katz, et al. Adapting an immersive virtual reality system for rehabilitation. *The Journal of Visualization and Computer Animation*, 14(5):261–268, 2003.

[15] T. Li, X. Zhou, K. Brandstatter, D. Zhao, K. Wang, A. Rajendran, Z. Zhang, and I. Raicu. Zht: A light-weight reliable persistent dynamic scalable zero-hop distributed hash table. In *Parallel & Distributed Processing (IPDPS), 2013 IEEE 27th International Symposium on*, pages 775–787. IEEE, 2013.

[16] Z. Lv, C. Esteve, J. Chirivella, and P. Gagliardo. Clinical feedback and technology selection of game based dysphonic rehabilitation tool. In *9th International Conference on Pervasive Computing Technologies for Healthcare (PervasiveHealth2015)*. IEEE, 2015.

[17] Z. Lv, C. Esteve, J. Chirivella, and P. Gagliardo. A game based assistive tool for rehabilitation of dysphonic patients. In *Virtual and Augmented Assistive Technology (VAAT), 2015 3rd IEEE VR International Workshop on*, pages 9–14, March 2015.

[18] Z. Lv, L. Feng, H. Li, and S. Feng. Hand-free motion interaction on google glass. In *SIGGRAPH Asia 2014 Mobile Graphics and Interactive Applications*. ACM, 2014.

[19] Z. Lv, S. Feng, L. Feng, and H. Li. Extending touch-less interaction on vision based wearable device. In *Virtual Reality (VR), 2015 IEEE*, pages 231–232, March 2015.

[20] Z. Lv, A. Halawani, S. Feng, S. ur Réhman, and H. Li. Touch-less interactive augmented reality game on vision-based wearable device. *Personal and Ubiquitous Computing*, 19(3-4):551–567, 2015.

[21] Z. Lv, A. Tek, F. Da Silva, C. Empereur-Mot, M. Chavent, and M. Baaden. Game on, science-how video game technology may help biologists tackle visualization challenges. *PloS one*, 8(3):57990, 2013.

[22] M. Munih, R. Riener, G. Colombo, L. Lünenburger, F. Müller, M. Slater, and M. Mihelj. Mimics: Multimodal immersive motion rehabilitation of upper and lower extremities by exploiting biocooperation principles. In *Rehabilitation Robotics, 2009. ICORR 2009. IEEE International Conference on*, pages 127–132. IEEE, 2009.

[23] S. Neil. Vireio. http://www.mtbs3d.com/new-vireio-site.

[24] M. M. North, S. M. North, and J. R. Coble. Virtual reality therapy: An effective treatment for psychological disorders. *Studies in health technology and informatics*, pages 59–70, 1997.

[25] S. Ottosson. Virtual reality in the product development process. *Journal of Engineering Design*, 13(2):159–172, 2002.

[26] G. PS, G. LJ, B. A, and B. E. Hyperbaric oxygen therapy. *JAMA*, 263(16):2216–2220, 1990.

[27] I. R. Hyperbaric oxygen therapy. *JAMA*, 264(14):1811, 1990.

[28] G. Riva. Virtual reality in psychotherapy: review. *Cyberpsychology & behavior*, 8(3):220–230, 2005.

[29] K. R. Scherer. What are emotions? and how can they be measured? *Social science information*, 44(4):695–729, 2005.

[30] P. M. Tibbles and J. S. Edelsberg. Hyperbaric-oxygen therapy. *New England Journal of Medicine*, 334(25):1642–1648, 1996. PMID: 8628361.

[31] TriDef. Tridef 3d. https://www.tridef.com.

[32] K. Wang, X. Zhou, H. Chen, M. Lang, and I. Raicu. Next generation job management systems for extreme-scale ensemble computing. In *Proceedings of the 23rd international symposium on High-performance parallel and distributed computing*, pages 111–114. ACM, 2014.

[33] Y. Wang, Y. Su, and G. Agrawal. A novel approach for approximate aggregations over arrays. In *Proceedings of the 27th International Conference on Scientific and Statistical Database Management*, page 4. ACM, 2015.

[34] S. Xavier. Trinus vr. http://trinusvr.com.

[35] J. Yang, B. Chen, J. Zhou, and Z. Lv. A low-power and portable biomedical device for respiratory monitoring with a stable power source. *Sensors*, 15(8):19618–19632, 2015.

[36] S. Zhang, D. Caragea, and X. Ou. An empirical study on using the national vulnerability database to predict software vulnerabilities. In *Database and Expert Systems Applications*, pages 217–231. Springer Berlin Heidelberg, 2011.

[37] S. Zhang, X. Zhang, and X. Ou. After we knew it: empirical study and modeling of cost-effectiveness of exploiting prevalent known vulnerabilities across iaas cloud. In *Proceedings of the 9th ACM symposium on Information, computer and communications security*, pages 317–328. ACM, 2014.

[38] X. Zhang, Z. Xu, C. Henriquez, and S. Ferrari. Spike-based indirect training of a spiking neural network-controlled virtual insect. In *Decision and Control (CDC), 2013 IEEE 52nd Annual Conference on*, pages 6798–6805. IEEE, 2013.

[39] D. Zhao, Z. Zhang, X. Zhou, T. Li, K. Wang, D. Kimpe, P. Carns, R. Ross, and I. Raicu. Fusionfs: Toward supporting data-intensive scientific applications on extreme-scale high-performance computing systems. In *Big Data (Big Data), 2014 IEEE International Conference on*, pages 61–70. IEEE, 2014.